\documentclass{emulateapj}
\def\stacksymbols #1#2#3#4{\def\theguybelow{#2}
        \def\verticalposition{\lower#3pt}
        \def\spacingwithinsymbol{\baselineskip0pt\lineskip#4pt}
        \mathrel{\mathpalette\intermediary#1}}
\def\intermediary #1#2{\verticalposition\vbox{\spacingwithinsymbol
        \everycr={}\tabskip0pt
        \halign{$\mathsurround0pt#1\hfil##\hfil$\crcr#2\crcr
                \theguybelow\crcr}}}
\def\lta{\stacksymbols{<}{\sim}{2.5}{.2}}
\def\gta{\stacksymbols{>}{\sim}{3}{.5}}

\shorttitle{IR Scatter in Early-Type  Galaxies}
\shortauthors{Mathews et al.}

\begin{document}

\title{VARIATIONS OF MID AND FAR-IR LUMINOSITIES AMONG 
EARLY-TYPE GALAXIES:\\
RELATION TO STELLAR METALLICITY AND COLD DUST}

\author{William G. Mathews\altaffilmark{1},
Pasquale Temi\altaffilmark{2}, Fabrizio Brighenti\altaffilmark{3,1}, Alexandre Amblard\altaffilmark{2} }
%\email{ptemi@mail.arc.nasa.gov}
%\email{mathews@ucolick.org}
%\email{fabrizio.brighenti@unibo.it}
%\altaffiltext{2}{
%Department of Physics and Astronomy, University of Western
%Ontario,
%London, ON N6A 3K7, Canada. }
\altaffiltext{1}{University of California Observatories/Lick
  Observatory,
Board of Studies in Astronomy and Astrophysics,
University of California, Santa Cruz, CA 95064
(mathews@ucolick.org).}
\altaffiltext{2}{Astrophysics Branch, NASA/Ames Research Center, MS
  245-6,
Moffett Field, CA 94035 (pasquale.temi@nasa.gov).}
\altaffiltext{3}{Dipartimento di Fisica e Astronomia,
Universit\`a di Bologna, via Ranzani 1, Bologna 40127, Italy
(fabrizio.brighenti@unibo.it).}

\begin{abstract}
The Hubble morphological sequence from early to late galaxies
corresponds to an increasing rate of specific star formation.  The
Hubble sequence also follows a banana-shaped correlation between 24
and 70 micron luminosities, both normalized with the K-band
luminosity.  We show that this correlation is significantly tightened
if galaxies with central AGN emission are removed, but the cosmic
scatter of elliptical galaxies in both 24 and 70 micron luminosities
remains significant along the correlation.  We find that the 24 micron
variation among ellipticals correlates with stellar metallicity,
reflecting emission from hot dust in winds from asymptotic
giant branch stars of varying metallicity.  Infrared surface
brightness variations in elliptical galaxies indicate that the K - 24
color profile is U-shaped for reasons that are unclear.  In some
elliptical galaxies cold interstellar dust emitting at 70 and 160
microns may arise from recent gas-rich mergers.  However, we argue
that most of the large range of 70 micron luminosity in elliptical
galaxies is due to dust transported from galactic cores by feedback
events in (currently IR-quiet) active galactic nuclei.  
Cooler dusty gas naturally accumulates in the cores of elliptical
galaxies due to dust-cooled local stellar mass loss and may  
accrete onto the central black hole, releasing energy.
AGN-heated gas
can transport dust in cores 
5-10 kpc out into the hot gas atmospheres
where it radiates extended 70 micron emission but is 
eventually destroyed by sputtering. 
This, and some modest star formation, defines a cycle of 
dust creation and destruction.
Elliptical galaxies evidently 
undergo large transient excursions in the banana plot in times 
comparable to the sputtering time or AGN duty cycle, 10 Myrs.  
Normally regarded as passive, 
elliptical galaxies are the most active galaxies 
in the IR color-color correlation.
\end{abstract}

\section{Introduction}

In a series of recent papers 
we describe observations and interpretations 
of far infrared emission from 
dust in early-type galaxies, 
emphasizing emission from dust that 
is in thermal contact with hot gas virialized 
in galactic or group potential wells
(Temi, Brighenti \& Mathews  
2005; 2007a,b; 2008; 2009a,b = TBM05; TBM07,a,b; 
TBM08; TBM09a,b).
We propose that interstellar dust is ejected 
and stripped from 
asymptotic giant branch (AGB) stars 
as they orbit through the hot interstellar gas 
($T \approx 10^6-10^7$K).
Interstellar dust is heated both by diffuse  
galactic starlight and by impacts of thermal electrons.
The loss of thermal energy by electron-grain impacts 
can cool gas 
near black holes in the galactic cores, 
possibly stimulating energetic feedback events 
(Mathews \& Brighenti 2003).
Dust both influences and is influenced by AGN events. 
In addition, mid infrared 
circumstellar emission observed 
in elliptical galaxies is emitted 
by hotter dust associated with outflows from 
mass-losing AGB stars.

When mid (24$\mu$m) and far (70$\mu$m) infrared 
luminosities from galaxies of all 
types are compared and normalized by the $K$-band luminosity,
$L_{24}/L_K$ and $L_{70}/L_K$,
galaxies 
of all Hubble types occupy a remarkably well-defined 
banana-shaped correlation.
Remarkably, the galactic Hubble sequence 
is ordered along the correlation from 
early to late morphologies (TBM09b). 
For spiral and irregular galaxies 
this color-color correlation
results from increasing 
specific star formation toward later type galaxies.
Evidently, dust heated to a range of temperatures by young stars 
emits mid and far infrared radiation 
in nearly the same proportion,
independently of the specific star formation rate. 

By contrast, early-type elliptical galaxies, 
in which star formation is small or absent, 
are contiguous with the earliest spiral galaxies 
in the color-color plot and occupy 
an extended region of decreasing $\log(L_{70}/L_K)$
with nearly constant 
$\log(L_{24}/L_K) \approx 30.2$.
In general, mid and far infrared emission from 
dust grains in early-type galaxies 
does not indicate the star formation rate.

Here we discuss 
an improved, sharper version of the banana color-color plot 
for normal galaxies 
in which galaxies with concentrated 
mid infrared emission (probably mostly AGNs) are removed.
In addition, we propose possible explanations for 
the cosmic scatter among elliptical galaxies in the 
infrared color-color plot.

\section{The Banana Plot}

The top panel of Figure 1, taken from TMB09b,
shows the banana-shaped  
($L_{24}/L_K$)-($L_{70}/L_K$)
plot for nearby galaxies observed in the 
Spitzer Infrared Nearby Galaxies Survey
(SINGS; Kennicutt et al. 2003).
In Figure 1, and subsequently,  
$L_{24}$ and $L_{70}$ are in 
cgs units and $L_K$ is in solar units.
The infrared Spitzer band luminosities at 24 and 70$\mu$m
are defined by $L_{\lambda} = \lambda F_{\lambda} 4 \pi D^2$
where $D$ is the distance.
The SINGS sample was selected to represent 
nearby normal galaxies of all morphological types.
The banana-shaped plot in the top panel of Figure 1 
provides an infrared main sequence 
for the Hubble classification.
Colors in Figure 1 (top) designate 
five morphological types using the de Vaucouleurs T
parameter from HyperLeda (Paturel et al. 2003).
The morphological type varies rather smoothly along the 
banana from early types at the lower left to the 
latest galaxies at the upper right: 
E $\rightarrow$ Sa,Sab $\rightarrow$ Sb,Scd $\rightarrow$ Irr. 
Proceeding upward in Figure 1,
% along the SINGS galaxies, 
the luminosity ratio 
$L_{24}/L_K \propto (L_{70}/L_K)^{1.2}$ 
is a surrogate for an increasing   
specific star formation rate (SFR) from early to late types 
(Calzetti et al. 2007). 
S0 galaxies have SFRs that vary from essentially zero, 
similar to  
red and dead E galaxies, to rates typical of 
irregular galaxies having the highest 
specific SFRs (TBM09a), 
i.e. S0s can be found all along the banana.

The galaxy sample discussed by TBM09b, 
selected from the Spitzer archive, 
is rich in early type galaxies, complementing 
the emphasis on later types in the SINGS data.  
The central panel of Figure 1, which combines 
these two data sets, shows that the lower region of 
the banana occupied by E and E-S0 galaxies is 
quite extensive and possible reasons for 
this extension are discussed below.

Recently we discovered 
that many of the outlying galaxies in the TBM09b sample 
displaced above 
the banana (indicated with green squares in Figure 1)
exhibit strongly concentrated 24$\mu$m emission.
This concentration 
is apparent from the visibility of Airy rings 
in 24$\mu$m images,  
including additional angular artifacts characteristic of  
MIPS diffraction at this wavelength 
(see also Calzetti et al. 2010). 
While it is possible that 
24$\mu$m diffraction in some galaxies  
may be due to centrally concentrated starbursts, 
most of the galaxies with Airy rings 
have been previously identified as having additional 
independent evidence of AGN activity. 
Moreover, in the central panel of Figure 1 
it is seen that galaxies having Airy rings (green squares) 
lie above the banana correlation 
{\it almost everywhere along the banana.}
If concentrated starbursts are responsible for Airy rings,
the star formation trajectory in Figure 1 of 
dust heated in unresolved starbursts would need to be 
{\it steeper} than the locus of star formation 
along the upper banana, 
$\delta\log (L_{24}/L_K)/\delta\log (L_{70}/L_K) \approx 1.2$,
which seems unlikely.
For these reasons, we argue that harder AGN radiation, 
not concentrated starbursts, is the most likely explanation for 
the 24$\mu$m Airy rings.
All early-type galaxies 
with Airy diffraction probably have low-level active nuclei 
(Tang et al. 2009) 

The top row of three panels in Figure 2 illustrates
24$\mu$m images of two galaxies with visible Airy rings,
NGC 1386 and NGC 0315,
and one, NGC 4472, which
has no detectable Airy diffraction. 
Galaxies in Figure 1 with significant visible Airy rings  
are listed in Table 1.
Centrally normalized surface brightness profiles 
are shown in the lower panel of Figure 2 
together with 
the Spitzer point response function at 24$\mu$m, 
which decreases 
to half its central value at about 3.4 arcseconds.
When (AGN) galaxies with strong 24$\mu$m Airy rings 
are removed, as in the bottom panel of Figure 1,
the infrared color-color banana plot becomes much tighter. 
%The one remaining upward outlier in Figure 1 
%(at $\log L_{24}/L_K = 0000$,$\log L_{70}/L_K = 0000$)
%is NGC 0000.
%[[must find out about this galaxy!]]

\subsection{Bottom of the Banana}

Figure 3 shows an expanded view of galaxies that occupy 
the bottom, rather flat part of the banana, 
most or all of which are E or E-S0 galaxies 
($T < -2.7$).
Galaxies with strong {\bf central} point source emission at 24$\mu$m 
have been removed.
Of particular interest is the large extent 
of the region occupied by early type galaxies, 
having an intrinsic scatter that 
exceeds the errors of observation 
(e.g. as shown in the vertical error bars in Fig. 4).
If these elliptical galaxies are dominated by 
old stellar populations, one might imagine 
that they would occupy a much smaller region in the 
banana plot.
We now discuss in turn 
the physical origin of the 
vertical ($L_{24}/L_K$) and horizontal ($L_{70}/L_K$)  
scatter of early-type galaxies in Figure 3.
The outlying galaxy with the lowest $L_{24}/L_K$ 
in Figures 1 and 3 is NGC 4406.
%[[What are the NGC names of the four lowest 
%outlying E galaxies 
%in Fig.3? If correct, 
%very low L24/LK could indicate an extremely 
%low [Z/H] (unlikely), 
%some other source of K emission, etc.]]

\section{Intrinsic Scatter of $L_{24}/L_K$ for E 
Galaxies: Stellar Metallicity}

Low mass, oxygen-rich asymptotic giant branch (AGB) stars are 
thought to dominate the optical luminosity and 
dust production in elliptical galaxies
(e.g. Habing 1996; Athey et al. 2002). 
Consequently, it is natural to expect that AGB stars 
having higher metallicity also have more 
hot circumstellar dust and more emission 
at 24$\mu$m, possibly 
accounting for the vertical scatter in Figure 3.
To test this idea, we use the sample of elliptical galaxies 
with metallicities $[Z/H]$ determined 
%{\bf (within $R_e/2$) } 
by Trager et al. (2000a, b), most of which have been observed 
at 24$\mu$m and which are 
identified with red circles in Figure 3. 
Relevant properties of Trager's sample are listed in Table 2.

Figure 4 shows that $L_{24}/L_K$ 
does indeed correlate 
with stellar metallicity 
$[Z/H] = \log[(Z/H)_{galaxy}/(Z/H)_{\odot}]$ is
normalized to the solar abundance: 
%\vskip.1cm
\begin{equation}
\log(L_{24}/L_K) = 29.99 \pm 0.31
+ (0.52 \pm 0.16)[Z/H]
\end{equation}
\vskip.1cm
\begin{equation}
\log(L_{24}/L_K) = 30.12 \pm 0.42  
+ (0.76 \pm 0.21)[Z/H]
\end{equation}
\vskip.1cm
%where 
%$Z$ is the fraction by number of elements 
%heaver than helium and 
Values of $[Z/H]$ in equations (1) and (2) 
refer respectively to apertures of size 
$R_e/8$ and $R_e/2$ where $R_e$ is the effective 
radius in the $B$-band (Trager et al. 2000b). 
The correlations in Figure 4 indicate that 
broadband observations 
at rest-frame 24$\mu$m could be used to determine 
the stellar metallicity of distant,
old-population early-type galaxies 
at higher redshift, 
provided there are no strong AGN contributions. 
The correlation between stellar age and metallicity 
among elliptical galaxies discussed by
Terlevich \& Forbes (2002),
is not apparent in the galaxies plotted
in Figure 4.
%We also note that the outlying galaxy 
%with the largest $\log L_{24}/L_K$ in Figure 4  
%is not an outlier in Figure 1.

\subsection{Metallicity Gradients and 24$\mu$m Surface Brightness}

Proceeding further with this idea, can the 
(negative) stellar metallicity gradients 
in elliptical galaxies be measured from
radial variations in the $K-24\mu$m color? 
%$L_{24}/L_K$?
An initial response to this question appears in
Temi, Brighenti, Mathews (2008 = TBM08),
and we follow the same data reduction procedure here.
In TBM08 we discussed the mean 
$\langle K-24\rangle$ color profile averaged over a sample 
of only 19 galaxies, 10 of which are in common
with the Trager et al. (2000b) sample.
In that paper we found that $\langle K-24\rangle(R)$
decreased linearly with $[Z/H]$ 
to about $R \approx 1.2R_e(K)$, 
then flattened, 
and at $R \gta 1.8R_e(K)$  began to rise again.
$R_e(K)$ is the K-band half-light radius.
Now with the complete 
Trager sample, about twice as large, we 
find a similar mean $\langle K-24\rangle$ differential 
surface brightness profile shown in 
the upper panel of Figure 5. 

The flattening of $\langle K-24\rangle$ 
at $0.8 \lta R/R_e(K) \lta 1.6$
and subsequent rise at larger $R/R_e(K)$ in Figure 5
does not indicate that these elliptical galaxies 
have bright halos of 24$\mu$m emission.
Figure 1 of TBM08 shows that the surface 
brightness profiles of typical elliptical galaxies 
at all Spitzer infrared wavelengths (including 24$\mu$m)  
deviate rather little from the de Vaucouleurs 
profiles in optical light.
But systematic radial color variations do exist and 
$\langle K-24\rangle$ is the most unusual. 
In the lower panel of Figure 5 we 
convert $\langle K-24\rangle$ into $[Z/H]$ using  
the linear relation shown in Figure 4 with $[Z/H]$ 
evaluated in the $R_e/2$ aperture.
We assume that the linear behavior in equation 2
can be extrapolated to larger $[Z/H]$. 
Values of $[Z/H]$ at $R_e/2$ are preferred to match 
the global luminosities $L_{24}$ and $L_K$ in Figure 4 
and because $R > R_e(K)$ for all $Z/H$ values in Figure 5.
When $[Z/H]$ values observed at $R_e/2$ are used, 
the U-shape of the radial variation in 
the lower panel of Figure 5 is largely 
unchanged but it is normalized to somewhat larger 
values of $[Z/H]$.

Optical observations of $[Z/H]$ 
profiles using the Lick indices 
at comparable $R/R_e(K)$ invariably indicate 
monotonically decreasing stellar metallicity 
with galactic radius 
and this trend might also be expected in the hot dust emission. 
But how can the flattening and possible rise 
in the $[Z/H]$ profile in Figure 5 be explained?
If the outer regions 
of elliptical galaxies are formed by the accretion 
and disruption  
of smaller galaxies, then we would expect 
lower metallicity at large $R/R_e(K)$  
typical of smaller galaxies, 
apparently opposite to the trend in Figure 5.
Excess $\langle K-24\rangle$ at large $R/R_e(K)$
could in principle be due to dust acquired by accretion
in distant regions of the galactic atmosphere
where the gas density 
(and possibly temperature) is much lower,
increasing the grain sputtering time, 
preserving a higher dust density.
However, in normal elliptical galaxies 
emission at 24$\mu$m 
is dominated by hot dust in
circumstellar regions around mass-losing AGB stars, 
not by emission from  
much colder interstellar dust grains.
(see for example our computed models in Fig. 8).

Does dust in low metallicity AGB winds 
have higher 24$\mu$m luminosities?
Computed evolutionary tracks of AGB stars 
exhibit increasing effective temperature 
with decreasing abundance 
(Cristallo et al. 2009).
If 24$\mu$m-emitting grains are heated by thermal impacts, 
this increase in effective temperature 
may be sufficient to account for the 
enhanced circumstellar 
24$\mu$m emission at large $R/R_e(K)$
seen in Figure 5. 
Another possible explanation is that the outflow 
velocity in AGB stars is 
reduced (or rendered intermittent) 
by lowered radiation pressure acting on fewer 
or smaller grains in low-metallicity AGB stars,
increasing the gas density in the AGB outflow, 
possibly  
allowing a smaller mass fraction of dust grains to 
emit disproportionally more in the 
mid-infrared. 
If low metallicity is the source of the 
apparent $[Z/H]$ upturn in Figure 5,
it must occur at metallicities lower than those 
plotted in Figure 4 where no low-$[Z/H]$ upturn is visible.

In any case, the decreasing  
$\langle K-24\rangle$ at small radii, 
$R/R_e(K) \lta 0.8$, 
in the upper panel of Figure 5 is consistent with 
the expected decrease in stellar abundance 
observed optically in this region.
The stellar metallicity gradient 
in Figure 5 at $R/R_e(K) \lta 0.8$,
$\Delta [Z/H]/\Delta \log R \approx -0.23$, 
is in good agreement with previously observed 
metallicity gradients in elliptical galaxies
(Carollo et al. 1993; Davies et al. 1993;
Kobayashi \& Arimoto 1999; Mehlert et al. 2003;
Kobayashi 2004; Koleva et al. 2011).
%This result, a measure of the 
%sensitivity of the $\langle K-24\rangle$ color 
%to decreasing stellar metallicity 
%in bright galactic cores, is consistent with the 
%correlations in equations 1 and 2 based on the 
%ratio of total galactic luminosities $L_{24}/L_K$.

\section{Intrinsic Scatter of $L_{70}/L_K$: 
Cold Interstellar Dust}

In the upper part of the banana, 
where $\log L_{70}/L_K \gta 31.1$ in Figure 1, 
24 and 70 micron emission from normal 
galaxies are dominated by star formation.
For $\log L_{70}/L_K \lta 31.1$, 
the 70 micron emission from early-type, mostly E galaxies 
is a measure of the varying amounts of cold 
dust in the hot interstellar gas (TBM07a,b).
 
To discuss horizontal variations and excursions 
in $L_{70}/L_K$ 
along the bottom of the banana, 
we return to the complete galaxy sample shown 
in the lower panel of Figure 1 
and consider only elliptical galaxies. 
Figure 6 shows the huge range in 
far infrared luminosity $L_{70}$ for elliptical 
galaxies ($T < -3.5$) having similar $L_K$.
Galaxies with the largest $L_{70}$ have  
either undergone dust-rich mergers (such as NGC 5018) 
or are central galaxies in groups, 
although some galaxies with the lowest $L_{70}$, 
such as NGC 1399, are also group-centered.
%The horizontal green dashed line marks the approximate 
%lower envelope to the $L_{70}$ distribution 
%in Figure 6. 
%*******from herschel11 sept1.tex
$L_{70}$ luminosities of 
NGC 1399 and 4472, 
both near the Spitzer detection limit, are
in excellent agreement with our 
steady state model for the minimum far infrared 
emission from interstellar dust ejected from AGB stars 
into the hot ISM of normal elliptical galaxies 
as described by TBM07a,b. 
In this steady state calculation, which we now briefly review, 
stellar dust is created locally at the same rate that it is 
consumed by sputtering.

Approximate stellar ages and main sequence turnoff rates
in elliptical galaxies can be obtained from
Balmer line absorption in optical spectra and photometry.
From this follows (1) the mean current stellar mass loss rate 
from old stellar populations,
$dM_*/dt \approx 1.5(M_*/10^{12} M_{\odot})$ 
$M_{\odot}$ yr$^{-1}$,
and (2) the 
associated interstellar dust production rate by AGB stars,  
assuming a metallicity-dependent
(dust mass)/(gas mass) ratio 
normalized to $\sim0.01$ at solar abundance.
%Warm ($T \sim 10^4$K), dusty gas is deposited into the hot
%($T \sim 10^6-10^7$K) interstellar gas  
%along the ram-pressure stripped wakes of orbiting AGB stars.
%Ejected dust is 
%heated (in part) by diffuse galactic starlight, 
%not the parent AGB star
%and maintained at $T \sim 10^4$ K 
%by UV from post-AGB stars.
Warm, dusty gas is deposited into the hot ($T \sim 10^6-10^7$K)
interstellar gas along the ram-pressure stripped wakes of orbiting AGB
stars and is subsequently maintained at $T \sim 10^4$K by UV from
post-AGB stars.
The stripped warm gas must thermally merge 
into the hot interstellar gas ($T \sim10^6-10^7$ K) in
less than about $10^{5.5}$ yrs, 
otherwise optical H$\alpha$ emission line luminosities 
would exceed typical observations (Mathews 1990; Mathews \& Brighenti 2003).

Consequently,
after $\sim10^{5.5}$ yrs dust grains can be expected to 
come into direct thermal contact with the
hot gas and begin to be eroded (sputtered) by collisions with
H$^{+}$ and He$^{++}$ ions.
A typical grain sputtering lifetime is  
$\sim10^7$ yrs at galactic radius $r \sim 5-10$ kpc.
Interstellar 
dust grains are heated in comparable amounts by absorption of
local starlight
and by inelastic collisions with thermal electrons in the hot gas.
Each colliding electron deposits $\sim(3/2)kT$ into the grain.

We assume that dust ejected from AGB stars,
has a typical initial grain size distribution
$N_0(r,a) \propto a^{-3.5}$ 
normalized to the local stellar
metallicity at galactic radius $r$.
In a steady state, when grain heating-cooling and
formation-destruction balance at every galactic radius $r$,
we can determine the temperature $T(r,a)$ of
(amorphous silicate) grains of radius $a$.
Integrating over the grain size distribution
(altered by sputtering) and the projected surface of the
galaxy,
we can calculate (TBM07a,b) the total specific luminosity
or flux $\lambda F_{\lambda}$ in any field of view aperture.
By this means the infrared SED can be estimated for 
any elliptical galaxy based on its luminosity, 
distance and the temperature and density profiles 
of hot interstellar gas observed in the X-ray. 

We find that the far infrared luminosity observed  
in certain galaxies -- such as NGC 1399 and 
NGC 4472 -- are very close to our steady state predictions 
(TBM07a). 
These galaxies are also near the lower limit of 
elliptical galaxies that can be detected 
with Spitzer MIPS at 70$\mu$m.
%near the dashed green line in Figure 6.
For these galaxies, the local hot 
interstellar gas is assumed to consume dust by sputtering 
at the same rate that it is produced in local AGB outflows.
This model can reproduce 70$\mu$m and 160$\mu$m luminosities 
for ellipticals with rather low $L_{70}$, 
but truly interstellar dust is too cold to emit 
observable 24$\mu$m emission.  
Evidently, $L_{24}$ is produced exclusively by much 
hotter dust in AGB outflows, 
heated only by the nearby parent AGB stars, 
and this emission is not considered in our dust evolution model.

If interstellar $L_{70}$ emission from all
elliptical galaxies agreed with our steady state predictions,
they would occupy a disconnected
island near $\log L_{70}/L_K \sim 29.0 \pm 0.5$.
And observed AGB-heated dust would still dominate 
24$\mu$m emission from ellipticals 
with some scatter for variable metallicity,
$\log L_{24}/L_K \sim 30.10 \pm 0.15$.
Our steady state model predicts that 
the banana plot would appear only as
this island for ellipticals and a separate powerlaw 
sequence of star-forming galaxies of later types.

However, and to our initial surprise, 
many optically normal E galaxies
-- e.g. NGC 4636 and 5044  -- 
have $L_{70}$ that are 10-30 times brighter than 
our steady state prediction.
This infrared excess indicates the presence of excess
cold interstellar dust 
(of mass $\sim 10^5$ $M_{\odot}$) that cannot
have come from local mass-losing stars.
In addition, elliptical galaxies with excess $L_{70}$ often have
spatially extended far infrared emission,
extending beyond the Spitzer 70$\mu$m point response 
function (TBM07a). 
Galaxies with observationally confirmed extended 
dust are indicated in Figure 6 with large red circles.
The far infrared color 
$L_{160}/L_{70}$
of observed excess dust emission 
confirms that this excess dust is 
located further out in the
E galaxy interstellar atmosphere, 
at $r\sim 5-10$ kpc, where grain heating
by starlight and electron-grain collisions 
are diminished (TBM07b).
Because of the short dust sputtering lifetime,
$\sim 10^7$ yrs, the excess dust emission at 
these large radii must be transient. 

Transient, spatially extended dust can arise (1) from 
recent mergers or (2) by internally produced dust that 
is hydrodynamically transported outward from galactic cores 
with AGN-heated gas. 
Some elliptical galaxies with the highest $L_{70}$ 
have acquired dust by a recent merger and sometimes 
a post-merger dust distribution is clearly visible. 
In other cases it is difficult to determine 
if dust is supplied externally for example 
by minor mergers since these may be 
difficult to detect without a
detailed analysis, as argued by Crockett et al. (2011).
Nevertheless, our impression from images of 
many elliptical galaxies with high $L_{70}$ is that they 
are not surrounded by small, gas-rich galaxies 
that are about to merge. 
If the dust sputtering lifetime $\sim10^7$ yrs is relevant,
such mergers would have to be quite frequent, 
requiring multiple simultaneous 
ongoing small mergers to explain for example  
the extended 70$\mu$m image of NGC 5044 (TMB07b). 
At a few kpc radius, the orbital time of merging galaxies 
is likely to be longer than the sputtering time 
of dust stripped from them.
In addition, gas-rich merging galaxies must 
release widespread dust into the hot gas atmosphere 
that emits 70$\mu$m, but star-forming regions in 
gas-rich merging galaxies must not emit detectable 24$\mu$m  
emission since emission at this  
wavelength correlates with the stellar metallicity 
of the host galaxy (Fig. 4).
Moreover, dwarf galaxies observed 
near central galaxies in groups/clusters 
tend to be free of interstellar gas, 
due in part to ram stripping in the distant past. 
For these reasons we consider here the hypothesis 
that internally produced interstellar dust can explain many 
elliptical galaxies with excess dust often extending 
far beyond the effective radius of the central galaxy.
In TBM07b we described how dusty, hot interstellar gas 
can be buoyantly 
or hydrodynamically transported outward 
as a result of AGN heating events 
(feedback) in galaxy cores.

For internally produced dust to
explain the extended excess infrared emission 
in elliptical galaxies, 
it is essential to have 
a reservoir of dust near central black holes.
This source of dust is conveniently provided by 
optically thick clouds or disks of dense dusty gas
observed in projection against starlight 
in galactic cores 
(e.g. van Dokkum \& Franx 1995; Lauer et al. 2005). 
The common incidence of dusty disks or chaotically distributed 
dusty clouds within the central kpc of many ellipticals 
indicates that the steady state dust creation-destruction 
assumption we use to compute extended far-IR dust emission  
breaks down in dense elliptical cores.
Apparently more dust is created in elliptical cores than can be 
destroyed by local sputtering. 
Dust in stellar ejecta can 
radiatively cool ambient 
hot gas faster than the sputtering or freefall time, 
as discussed by Mathews \& Brighenti (2003).
When the central black hole accretes some nearby 
dust-cooled gas and releases feedback energy, 
we propose that some of the central dusty gas 
is heated and flows buoyantly out to 5-10 kpc from the center 
where it thermally merges with the hot interstellar gas  
before the dust is sputtered away.
We identify these AGN-driven outflows 
as the source of excess extended far infrared emission 
observed in many elliptical galaxies 
shown with red circles in Figure 6.
%At the present time, after cold dust is 
%removed from galactic cores, these AGN no longer 
%emit concentrated nonthermal infrared emission 
%since these galaxies have been removed from the banana.

Figure 7 shows elliptical galaxies ($T < -3.5$)
in the bottom of the banana plot
in a region similar to Figure 3.
Galaxies with red dots in Figure 7 are chosen 
because of their large $L_{70}/L_K$
for given $L_{24}/L_K$, 
lying along the rightmost boundary of the lower banana.
These same galaxies are also plotted 
with red dots in Figure 6 where it is seen that they all 
have significant excess 70$\mu$m emission.
Galaxies with red circles in both Figure 6 and 7 
are those in which the excess far-IR emission 
is known to be spatially extended 
in 70$\mu$m Spitzer images (TBM07a,b). 

Small open green circles in 
Figures 6 and 7 show the approximate computed 
locations of two galaxies 
-- NGC 5044 and NGC 4636 -- 
using our steady-state dust evolution model 
in which dust ejected from galactic 
stars is destroyed locally by sputtering.
The green arrows 
illustrate the large additional (excess) contribution to 
$L_{70}$ which we interpret as 
dust in gas heated by AGN feedback events  
and then transported outward from galactic cores.
Without this extended excess dust these two galaxies 
may not have been detected with Spitzer.
Similar displacements probably occurred recently 
in the other red elliptical galaxies in the 
banana boundary region of Figure 7.

If our interpretation is correct,
group-centered galaxies, 
like NGC 5044 and NGC 4636, move back and forth 
along the bottom of the banana in Figure 1 on timescales 
determined by the grain sputtering lifetime and the interval 
between AGN outbursts, which may be roughly comparable, 
$\sim10^7$ yrs. 
Elliptical galaxies that are not known to be group-centered 
can also participate in similar large, transient
horizontal excursions in the banana plot, 
provided their virialized hot gas atmospheres 
are hot enough ($T \gta 4\times10^6$ K) 
to sputter and are sufficiently extended. 
However, there appears to be no correlation between 
excess 70$\mu$m emission and X-ray luminosity 
(TBM07a).

One potential difficulty with our 
hypothesis is that 
no galaxy is observed to lie 
appreciably further to the right than the 
group of red dot galaxies in Figure 7.
Comparing Figure 7 and the bottom panel of Figure 1,
it is seen that the slope of 
the rightmost envelope of red galaxies in 
Figure 7 lies just along the star-formation 
track defined by galaxies of later type 
($(L_{24}/L_K) \propto {(L_{70}/L_K)}^{1.2}$).
Somehow, the rightward excursions caused 
by the outward movement of excess 
cold dust due to feedback events 
must be limited so as not to exceed the star-formation locus.

To understand this, 
we conclude that central star formation must eventually occur 
when the mass of dusty gas accumulates 
beyond a certain point in the cores of 
elliptical galaxies. 
Star formation causes both $L_{70}$ and $L_{24}$ 
to increase as elliptical galaxies rise along the 
SINGS star-forming locus.
Star formation and AGN-driven dust transport into 
the hot gas atmospheres are not mutually exclusive.
If central star formation commences, 
the horizontal green arrows in Figure 7 
should bend at the right to move up along the 
star-formation trajectory with increasing $L_{24}/L_K$.
It is likely that low-level star formation occurs 
in all elliptical galaxy cores 
(Ford \& Bregman 2012) 
in dense dusty clouds formed by dust-assisted cooling 
of gas ejected from AGB stars 
(Mathews \& Brighenti 2003). 
%However, the absence of observed optically dark dust clouds 
%in some elliptical cores may correspond to an immediate 
%post-feedback period and does not necessarily imply 
%a long-lasting absence of central dust or star formation. 

A quasi-steady-state cyclic dust evolution  
that involves reservoirs of 
increasing dust mass in galaxy cores  
may explain the large variation in $L_{70}/L_K$ 
shown in Figures 6 and 7. 
Cold gas clouds containing optically visible dust 
slowly accumulate in galaxy cores as gas 
from stellar mass loss is cooled by dust.
Occasionally these dusty clouds 
may be heated by AGN events to $\sim 10^7$ K
and flow out into the surrounding hot atmosphere 
emitting 70$mu$m radiation.
Ultimately this extended dust is likely to   
be sputtered into the gas phase.  
Such a hypothetical dust evolution could 
create and destroy dust at approximately balancing rates.
Star formation may or may not occur during each 
feedback episode, and it would also serve to consume 
some of the accumulated cold gas and dust. 
%without resulting in star formation.

The evolution of optically thick dust clouds in the 
cores of elliptical galaxies must also participate 
in the quasi-cyclic production of extended cold dust.
If cold dusty gas within the central $\sim$kpc 
forms by cooled stellar ejecta as we propose, 
the default endstate is most likely a small 
rotating dusty disk. 
The spin axis of the disk will be aligned with 
the kinematical minor axis of nearby stars. 
While such central disks are often observed
(e.g. van Dokkum \& Franx 1995; Lauer et al. 2005), 
it is more common to find a multitude of patchy, 
chaotically distributed dust 
clouds or no dust absorption at all.
Because of the small solid angle 
presented to the central black hole, 
thin, cold dusty disks may be difficult to disrupt 
%and heat to sputtering temperatures 
%particularly 
if AGN energy 
is released only at the black hole. 
However, disks may be more easily disrupted by symmetric 
depositions of AGN energy in bipolar regions 
somewhat removed from the black hole and misaligned from the 
disk axis. 
In this case shock energy reaches opposite sides of the disk 
at different times, causing mechanical disk disruption.
Even if strong shock waves arrive simultaneously 
on both sides, if the shocks are strong enough, 
much or all of the disk could be sufficiently heated to merge 
with the hot gas. 
While these important details remain unexplored, 
in at least one case we observe both a relaxed dusty disk 
with nearby dense clouds presumably from a disrupted previous 
disk generation. 
In NGC 5044, where extended dust is resolved 
at 70$\mu$m (TMB07b), the central kpc contains many chaotically 
distributed optically thick clouds with orbital lifetimes 
$\sim10^7$ yrs compatible to the 
hydrodynamic and sputtering lifetimes 
of the extended dust (TBM07b). 
In many cases remnants of 
disrupted central dusty gas may be 
too small or too widely dispersed by AGN outbursts to be  
visible as optically thick clouds.

%\end{document}

\section{Un-normalized Banana Plot}

Figure 8 shows a less compact version of
the banana plot in which $L_{24}$ and $L_{70}$ are
plotted without normalization with $L_K$.
SINGS galaxies (crosses in Fig. 8) 
define a star-forming sequence
$L_{24} \propto {L_{70}}^{1.08}$
in which $L_{24}/L_{70}$ is approximately constant
as $L_{24}$ varies by five orders of magnitude. 
Many of the Spitzer-archived elliptical galaxies in Figure 8 
that occupy the SINGS star-forming region above
$\log L_{24} \approx 41.8$
are peculiar.\footnotemark[2]
SINGS galaxies in Figure 8 having low star formation rates
lie considerably below $L_{24}$ of normal giant elliptical 
galaxies.
The black filled squares in Figure 8 illustrate
the very small diffuse interstellar $L_{24}$ emission 
for five representative ellipticals 
computed with our steady-state dust evolution.

\footnotetext[2]{
%Some of the Spitzer-archived 
%E galaxies in Figure 8 that lie up along the SINGS 
%star-forming track at $\log L_{24} > 41.8$ 
%are peculiar. 
The two elliptical (red) galaxies with $\log L_{24} \sim 42.6$ 
(NGC 807 and UGC 01503) in Figure 8 have massive 
($\sim 10^9$ $M_{\odot}$) multi-kpc rotating dusty HI disks 
(Goudfrooij et al. 1994; Young et al. 2009; Young 2002), 
likely to be old mergers.
NGC 3125 (at $\log L_{24} \approx 42.26$) 
is a compact dwarf galaxy ($\sigma = 48$ km s$^{-1}$) 
with strong emission lines (Koprolin and Zeilinger 2000).
Very little is known about NGC 4786 
($\log L_{24} \approx 42.26$).
The group of five E galaxies at 
$\log L_{24} \approx 41.92$ includes NGC 5018 
[with a large $\sim 10^8$ $M_{\odot}$ filament of dusty HI gas],
NGC 3557 [an FRI radio source (Birkinshaw \& Davies 1985) with 
extended H$\alpha$ emission (Goudfrooij et al. 1994)],
NGC 4125 [with a central X-ray source (Flohic et al. 2006)],
IC 4296 [an FRI radio galaxy with optical line emission and 
a dusty core], 
and IC 1459 [which contains detectable HI(Doyle et al. 2005)].
Finally, very little is known 
about the the outlier S0 galaxy NGC 526 at 
$\log L_{24} \approx 43.45$,$\log L_{70} \approx 43.0$ 
in Figure 8.
}

\section{Conclusions}

The banana-shaped infrared color-color
correlation for normal galaxies --
a plot of $\log (L_{70}/L_K)$ vs. $\log (L_{24}/L_K)$
(bottom panel of Figure 1) --
approximately follows the morphological progression
along Hubble types.
As $\log (L_{24}/L_K)$ and $\log (L_{70}/L_K)$
increase with the specific star-forming rate,
the dominant Hubble type becomes progressively later:
Sa,Sab $\rightarrow$ Sb,Scd $\rightarrow$ Irr.
By contrast, most E galaxies 
lie near the bottom of the
banana in a contiguous region of restricted 
$\log (L_{24}/L_K)$ and decreasing $\log (L_{70}/L_K)$.
Some gas-rich E galaxies lie along the star-forming
locus of late type galaxies and S0 galaxies  
are found throughout the banana.
Our concern here is to understand the 
variation of $\log (L_{24}/L_K)$ and $\log (L_{70}/L_K)$ 
among normal elliptical galaxies in the banana plot.

We show that the banana-shaped correlation, first discussed 
by TBM09, becomes much tighter 
when galaxies with AGN emission are removed. 
Dust grains can be significantly heated by AGN radiation
and emit strongly at 24$\mu$m 
within several arcseconds of the galactic centers.
These galaxies, having hard, highly concentrated 
infrared emission,
can be identified by the presence of
Airy diffraction rings at 24$\mu$m.
AGN galaxies of all types with Airy rings 
lie above the infrared banana correlation.
To achieve this, centrally illuminated 
dust would need to move along a trajectory
$\delta\log (L_{24}/L_K)/\delta\log (L_{70}/L_K)$ 
that is steeper than the star formation 
trajectory along the upper banana,
$\delta\log (L_{24}/L_K)/\delta\log (L_{70}/L_K) \approx 1.2$.
Because of this slope difference, 
we suggest that the concentrated source 
of central IR radiation is due to AGN rather than 
small starburst regions unresolved at 24$\mu$m. 

An increasing specific star formation rate from early 
spiral to irregular galaxies with 
$L_{24} \propto L_{70}^{1.08}$
apparently explains the progressive 
transition toward later galactic Hubble types 
upward along the banana.
The bottom part of the infrared banana is populated 
by elliptical galaxies with a relatively 
small but significant scatter in 24$\mu$m emission, 
$\log(L_{24}/L_K) \approx 30.15 \pm 0.15$,
and a much larger scatter in 70$\mu$m emission,
$29 \lta \log(L_{70}/L_K) \lta 31$ 
in units of (erg s$^{-1}$)/$L_{K\odot}$.

We find that the scatter in 
$\log(L_{24}/L_K)$ among elliptical galaxies correlates 
with their stellar metallicity,
$(L_{24}/L_K) \propto [Z/H]^{0.7}$. 
Emission from hotter dust at 24$\mu$m 
is likely to originate in circumstellar regions 
near AGB stars where dust is thought to be produced.
The correlation of 24$\mu$m emission 
with metallicity is probably due to the simple expectation 
that outflows from metal-rich AGB stars 
produce more dust per gram of gas ejected. 

But the surface brightness variation in 24$\mu$m emission 
from elliptical galaxies, 
while largely mimicking the de Vaucouleurs profile, 
contains unexpected color variations when compared with  
the $K$-band.
When plotted against galactic radius normalized by
the $K$-band effective radius $R/R_e(K)$,
the mean $\langle K-24\rangle$ color  
for a sample of well-observed elliptical galaxies 
first decreases in a power-law fashion, 
then undergoes a broad minimum followed by a 
rise at large radius $R/R_e(K)$.
The initial decrease is consistent 
with decreasing stellar metallicity with 
radius similar to that found at optical wavelengths,
$\Delta [Z/H]/\Delta \log R \approx -0.3$.
The origin of the flattening and possible rise in 
$\langle K-24\rangle$ beyond $R/R_e(K) \sim 1$ 
is uncertain.
If hot dust in AGB winds is thermally heated, 
increasing $\langle K-24\rangle$ with decreasing stellar 
metallicity may be related to  
increasing effective temperature 
of AGB stars with decreasing metallicity.
Alternatively, 
lower velocity (radiation pressure-driven) outflows 
from metal poor AGB stars might have larger 24$\mu$m emissivity.

The intrinsic scatter of elliptical galaxies 
in the banana-shaped infrared correlation 
at longer wavelengths, 70$\mu$m and 160$\mu$m, 
is much larger.
Recent gas-rich mergers may be a source of cold dust 
for some elliptical galaxies. 
However, for most ellipticals
we propose that this scatter is due to transient 
hydrodynamic flows of cold dust from galactic cores 
far out into the surrounding  
hot galactic interstellar gas
following AGN feedback events. 
The variation of $\log(L_{70}/L_K)$ 
is uncorrelated with X-ray luminosity $L_X$ 
of the galactic/group baryonic halo.
The $L_{70}/L_K$ ratios of 
some group-centered E galaxies like 
NGC 4472 and 1399 are 
in good agreement with a steady state model in which 
dust created by AGB stars 
is locally destroyed by sputtering 
at each radius in the galactic atmospheres.
Other ellipticals 
such as NGC 4636 and 5044 have much larger 
$L_{70}$ luminosities from excess cold dust 
that cannot be produced by local stars.
Lacking convincing 
evidence of recent gas-rich mergers in these 
otherwise normal elliptical galaxies, 
we propose that AGN outbursts cause 
excess dust to be hydrodynamically 
transported out to 5-10 kpc into the hot 
virialized gaseous atmospheres. 
Spatially extended 70$\mu$m emission has been confirmed 
in relatively nearby elliptical galaxies (TBM07a,b). 
Evidently, values of $L_{70}/L_K$ for elliptical galaxies 
along the bottom of the banana plot are transient 
and may vary on $\sim10^7$ yr timescales, depending on 
the grain sputtering time and the AGN duty cycle.
The outward relocation of variable masses of cold dust 
from the central reservoir, observed at different 
phases of the outflow, may explain 
why galaxies of similar $L_K$ have $L_{70}$
luminosities that vary over two orders of magnitude.

However, the continuity of the banana plot from 
E galaxies to early spirals 
(where $L_{24}/L_K$ and $L_{70}/L_K$ 
are due to star formation) 
requires that the enhanced 70$\mu$m emission 
from extended excess cold 
dust in E galaxies 
must not exceed $L_{70}/L_K$
as produced by dust in star-forming galaxies of later types.
To explain this unlikely coincidence, 
we speculate that star formation naturally begins 
in the cores of elliptical galaxies 
as the mass of accumulated cold gas and dust 
approaches some star-forming threshold. 
This conjecture follows from the 
expected secular accumulation of dust-cooled gas 
from AGB stellar mass loss in the central $\sim$kpc  
of elliptical galaxies (Mathews \& Brighenti 2003).

\vskip.2cm
\acknowledgements
This work is based on observations made with the Spitzer
Space Telescope, which is operated by the Jet Propulsion
Laboratory, California Institute of Technology, under NASA
contract 1407. 
Studies of
the evolution of hot gas in elliptical galaxies at UC Santa Cruz
are supported by an NSF grant for
which we are very grateful. 
Support for this work was provided by NASA ADP Grant.
F.B. aknowledges financial support from PRIN MIUR 2010-2011, prot. 2010LY5N2T.
We acknowledge the usage of the HyperLeda database (http://leda.univ-lyon1.fr)
and the NASA/IPAC Extragalactic Database (NED).

%Since both colors are normalized with $L_K$,
%the $K$-band luminosity, and therefore the stellar mass,
%increases systematically
%from the upper right to the lower left in Figure 1.

%\end{document}

\clearpage

\begin{deluxetable}{lccccc}
\tabletypesize{\tiny}
\tablecaption{Galaxies in Figure 1 with Airy Diffraction\tablenotemark{a}}
\tablewidth{9.2cm}
\tablehead{
%\colhead{Name} & \colhead{Log $L_{Ks}$} & \colhead{Log$L_{24\mu m}$} &
%\colhead{Log$L_{70\mu m}$}\\
%\colhead{} & \colhead{($L_{Ks,\odot}$)} & 
%\colhead{$(erg \ s^{-1}$)} & \colhead{$(erg \ s^{-1}$)}\\
\colhead{Name} & \colhead{T-Type\tablenotemark{b}} & \colhead{D} &
\colhead{Log $L_{Ks}$} & \colhead{Log$L_{24\mu m}$} &
\colhead{Log$L_{70\mu m}$} \\
\colhead{} & \colhead{} & \colhead{(Mpc)} & \colhead{($L_{Ks,\odot}$)} &
\colhead{$(erg \ s^{-1}$)} & \colhead{$(erg \ s^{-1}$)}\\
}
\startdata
NGC~0315 & -4   & 58.9 & 11.71 & 42.68 & 42.77 \\
NGC~0814 & -1.8 & 20.7  & 9.508 & 42.65 & 42.82 \\
NGC~0526 & -2.0 & 78.7  & \nodata   & 43.44 &  43.02 \\
NGC~1266 & -2.1 & 30.1 & 10.51 & 43.07 & 43.77 \\
NGC~1386 & -0.8 & 16.5 & 10.56 & 42.69 & 42.98 \\
NGC~1377 & -2.1 & 22.2  & 10.15 & 43.13 & 43.2 \\
NGC~2110 & -3.0 & 31.3 & 11.08 & 42.97 & 43.4 \\
NGC~2325 & -4.7 & 31.9 & 11.2  & 41.62 & 41.3 \\
NGC~3226 & -4.8 & 23.5 & 10.66 & 41.35 & 41.97 \\
NGC~3265 & -4.8 & 21.3 & 9.83  & 42.3  & 42.8 \\
NGC~4261 & -4.8 & 31.6 & 11.44 & 41.88 & 41.81 \\
NGC~5077 & -4.8 & 32.4 & 11.08 & 41.72 & 41.85 \\
NGC~6776 & -4.1 & 70.4 & 11.43 & 42.08 & 42.5 \\
NGC~5273 & -1.9 & 16.5 & 10.32 & 41.53 & 41.96 \\
IC~~5063 & -0.9 & 45.3 & 11.16 & 43.82 & 43.67 \\
E103-035 & -0.3 & 54.2  & 10.68 & 43.87 & 43.43 \\
ESO428-014 & -1.7 & 21.3 & 10.5  & 42.99 & 43.04 
\enddata
\tablenotetext{a}{These galaxies are identified with green squares 
in the central panel of Figure 1.}
\tablenotetext{b}{
Morphological type from HyperLeda.}
\end{deluxetable}

\clearpage

\begin{deluxetable}{rccccccrcccc}
\tabletypesize{\tiny}
\tablecaption{Relevant Properties of the Trager et al. Sample}
\tablewidth{17.4cm}
\tablehead{
\colhead{Name} & T-Type\tablenotemark{a} & \colhead{Log $L_{Ks}$} 
& \colhead{Log$L_{24\mu m}$} &
\colhead{Log$L_{70\mu m}$}
& \colhead{[Z/H]\tablenotemark{b}} & \colhead{[Z/H]\tablenotemark{b}}
& \colhead{Age\tablenotemark{b}} 
& \colhead{D} & \colhead{$R_e(K)$} & \colhead{$R_e(K)$}
& \colhead{${R_{max}/R_e}\tablenotemark{c}$} \\
\colhead{(NGC)} & \colhead{} & \colhead{($L_{Ks,\odot}$)} & \colhead{$(erg s^{-1}$)} 
& \colhead{$(erg \ s^{-1}$)} &($R_e/8$) & ($R_e/2$)
& \colhead{(Gyr)} & \colhead{(Mpc)} &
\colhead{($\prime \prime$)} & \colhead{(kpc)} & \colhead{}\\
}
\startdata
221  & -4.7 &   9.13  &  39.22 &  37.93 &  0.00$\pm$0.05 & -0.08$\pm$ 0.05 &4.9$\pm$1.3 &
0.81     & 27.0 &  0.11      & 2.72 \\
315\tablenotemark{d}  & -4.0 &  11.71  &  42.68 &  42.77 &
0.32$\pm$0.06 & 0.17$\pm$0.06 &6.9$\pm$1.9 &
58.88    & 22.2 &  6.36      & 0.96 \\
507  & -3.3 &  11.68  &  42.02 &  \nodata &  0.18$\pm$0.06 & 0.28$\pm$0.13 &3.5$\pm$2.7 &
67.19    & 24.0 &  7.84      & 1.36 \\
584  & -4.6 &  11.18  &  41.48 &  41.18 &  0.48$\pm$0.03 & 0.26$\pm$0.07 &3.4$\pm$1.1 &
23.76    & 23.3 &  2.69    & 1.62 \\
636  & -4.8 &  10.67  &  40.73 &  \nodata &  0.34$\pm$0.07 &  0.10$\pm$0.05&6.8$\pm$1.4 &
22.28   & 15.3 &  1.66     & 1.39 \\
720  & -4.8 &  11.14  &  41.42 &  40.69 &  0.44$\pm$0.15 & 1.13$\pm$0.42 & 1.1$\pm$0.8 &
22.29   & 25.2 &  2.73      & 1.57 \\
821  & -4.8 &  10.95  &  41.06 &  \nodata &  0.22$\pm$0.03 & 0.12$\pm$0.05 & 7.1$\pm$1.2 &
24.09    & 21.9 &  2.56      & 1.07 \\
1374  & -4.3 &  10.68  &  40.73 &  \nodata &  0.13$\pm$0.07 &\nodata &  9.5$\pm$2.6 &
19.77    & 15.4 &  1.48      & 1.25 \\
1399  & -4.5 &  11.40  &  41.49 &  40.50 &  0.29$\pm$0.06 &\nodata &11.5$\pm$2.4 &
19.40     & 32.9 &  3.10      & 2.83 \\
1404  & -4.7 &  11.20  &  41.45 &  40.80 &  0.25$\pm$0.05 & \nodata& 9.0$\pm$2.5 &
19.40    & 18.7 &  1.76      & 3.57 \\
1427  & -4.0 &  10.66  &  40.90 &  40.89 & -0.07$\pm$0.03 & \nodata& 12.2$\pm$1.6 &
23.55    & 21.5 &  2.46      & 1.45 \\
1700  & -4.7 &  11.27  &  41.51 &  41.35 &  0.50$\pm$0.03 & 0.32$\pm$0.05& 2.8$\pm$0.5 &
38.04     & 14.9 & 2.76       & 1.57 \\
2300  & -3.4 &  11.18  &  41.36 &  \nodata &  0.36$\pm$0.04 &0.14$\pm$0.04 &10.1$\pm$2.0 &
27.67     & 22.3 &  2.99      & 1.54 \\
2778  & -4.7 &  10.26  &  40.37 &  \nodata &  0.29$\pm$0.07 &-0.04$\pm$0.06 & 14.9$\pm$3.5 &
22.91     & 10.5 & 1.17       & 1.67 \\
3377  & -4.8 &  10.47  &  40.50 &  40.74 &  0.19$\pm$0.06 & -0.12$\pm$0.04& 5.9$\pm$1.2 &
11.22    & 25.4 & 1.39       & 0.76 \\
3379  & -4.8 &  10.89  &  41.03 &  40.56 &  0.21$\pm$0.04 & 0.00$\pm$0.04 &13.2$\pm$2.4 &
10.57    & 28.5 &  1.46      & 1.09 \\
3608  & -4.8 &  10.83  &  41.00 &  41.27 &  0.26$\pm$0.04 & 0.07$\pm$0.07& 9.0$\pm$2.5 &
22.91   & 16.2 &   1.80     & 1.44 \\
4261\tablenotemark{d}  & -4.8 &  11.44  &  
41.88 &  41.81 &  0.18$\pm$0.04 &-0.01$\pm$0.03 & 21.0$\pm$1.7&
31.62    & 24.2 &  3.72      & 1.17 \\
4374  & -4.2 &  11.39  &  41.45 &  42.03 &  0.12$\pm$0.03 & -0.01$\pm$0.05&14.4$\pm$2.8 &
18.37     & 33.5 &   2.99     & 1.12 \\
4472  & -4.8 &  11.65  &  41.71 &  40.96 &  0.25$\pm$0.05 & 0.18$\pm$0.06& 8.4$\pm$2.7 &
17.06    & 56.1 &   4.65     & 0.89 \\
4478  & -4.8 &  10.52  &  40.74 &  \nodata &  0.29$\pm$0.10 & -0.01$\pm$0.04 & 10.3$\pm$1.7 &
18.11    & 11.8 &  1.04      & 1.35 \\
4489  & -4.8 &  10.10  &  40.11 &  \nodata &  0.14$\pm$0.06 & -0.15$\pm$0.04 & 4.6$\pm$0.5 &
17.86     & 14.1 & 1.22       & 1.13 \\
4552  & -4.6 &  11.03  &  41.26 &  41.06 &  0.27$\pm$0.04 & 0.10$\pm$0.04  &13.1$\pm$2.7 &
15.34     & 22.8 &   1.70     & 1.65 \\
4649  & -4.6 &  11.52  &  41.58 &  40.86 &  0.27$\pm$0.04 &0.05$\pm$0.04 &18.3$\pm$2.8 &
17.06    & 42.1 &  3.49      & 0.46 \\
4697  & -4.8 &  11.22  &  41.26 &  41.92 &  0.06$\pm$0.05 &-0.29$\pm$0.04 &15.6$\pm$2.5 &
16.22   & 39.5 &  3.11      & 0.65 \\
5813  & -4.8 &  11.40  &  41.37 &  41.51 & -0.04$\pm$0.03 & -0.20$\pm$0.06& 24.3$\pm$2.0 &
32.21    & 36.2 &  5.66      & 0.95 \\
5831  & -4.8 &  10.84  &  41.10 &  \nodata &  0.54$\pm$0.04 &0.18$\pm$0.06 & 4.2$\pm$1.0 &
27.16   & 19.4 &   2.56     & 0.75 \\
5846  & -4.7 &  11.36  &  41.43 &  41.53 &  0.14$\pm$0.05 &-0.13$\pm$0.04 &23.1$\pm$2.7 &
24.89     & 32.7 &  3.96      & 2.47 \\
6703  & -2.8 &  11.06  &  41.34 &  41.36 &  0.30$\pm$0.06 & 0.03$\pm$0.06& 7.9$\pm$2.1 &
32.06    & 20.1 &  3.13      & 1.16 \\
7619  & -4.7 &  11.58  &  41.62 &  \nodata &  0.20$\pm$0.03 & 0.07$\pm$0.04&15.1$\pm$2.2 &
52.97    & 16.4 &  4.22      & 1.64 \\
7626  & -4.8 &  11.34  &  41.44 &  41.17 &  0.16$\pm$0.04 &-0.06$\pm$0.02 &18.2$\pm$2.0 &
39.99     & 20.7 &  4.02      & 1.13 \\
7785  & -4.8 &  11.46  &  41.58 &  41.45 &  0.20$\pm$0.04 &-0.01$\pm$0.04 &15.0$\pm$2.3 &
55.46     & 17.2 &  4.64      & 0.84 \\
\enddata
\tablenotetext{a}{Morphological type from HyperLeda.}
\tablenotetext{b}{Values of $[Z/H]$ for $R_e/8$ and $R_e/2$ 
are ``model 4'' values from
Trager et al. (2000a,b) refer to the $B$-band 
effective radius $R_e$ which on average is about
1.45 times larger than the $K$-band value $R_e(K)$ (TBM08). Ages listed
above refer to the $R_e/2$ observations.}
\tablenotetext{c}{
Maximum extent of observable 24$\mu$m emission.
}
\tablenotetext{d}{Identified as an AGN galaxy from 24$\mu$m Airy rings.}

\end{deluxetable}

%\end{document}

\clearpage

\begin{figure}%1
%\vskip3.in
\centering
\includegraphics[width=7.in,scale=1.0,angle=0]{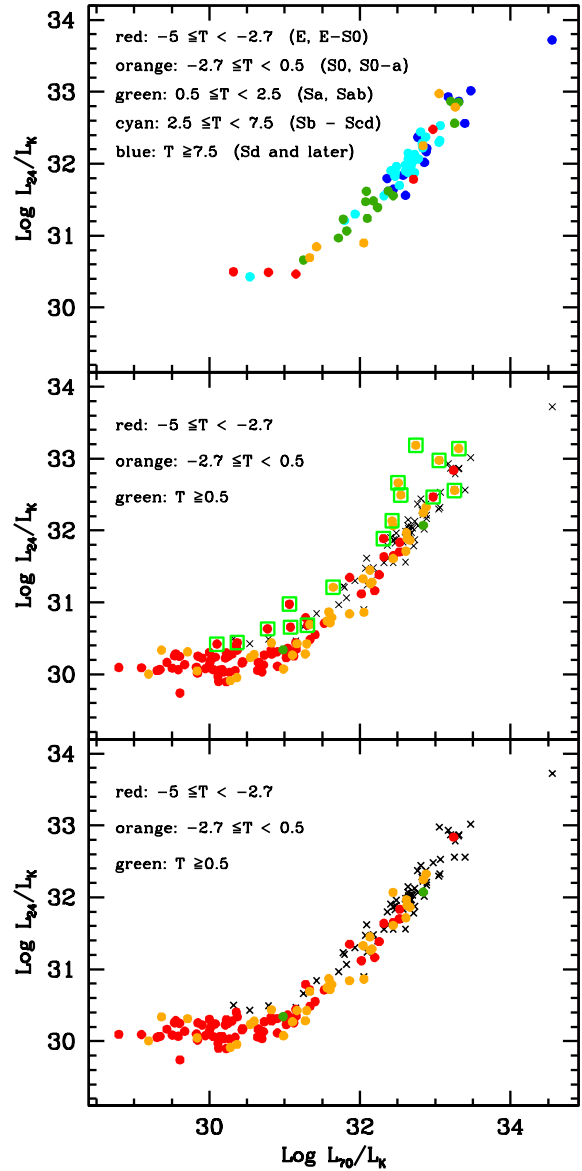}
%\vskip.7in
\caption{
Upper panel: Infrared color-color banana plot
$L_{24}/L_K$ vs. $L_{70}/L_K$ for
SINGS galaxies showing progression of Hubble types.
Central panel: IR color-color plot showing
galaxies in the TBM09b sample with X indicating
SINGS galaxies. Green squares indicate TBM09b
(AGN or starburst) galaxies with centrally concentrated
24$\mu$m emission.
Bottom panel: Same as central panel without galaxies
having centrally concentrated 24$\mu$m emission.
}
\label{fig1}
\end{figure}

\clearpage

\begin{figure}
\centering
%\vskip-1.0in
\includegraphics[width=3.9in,scale=1.0,angle=0]{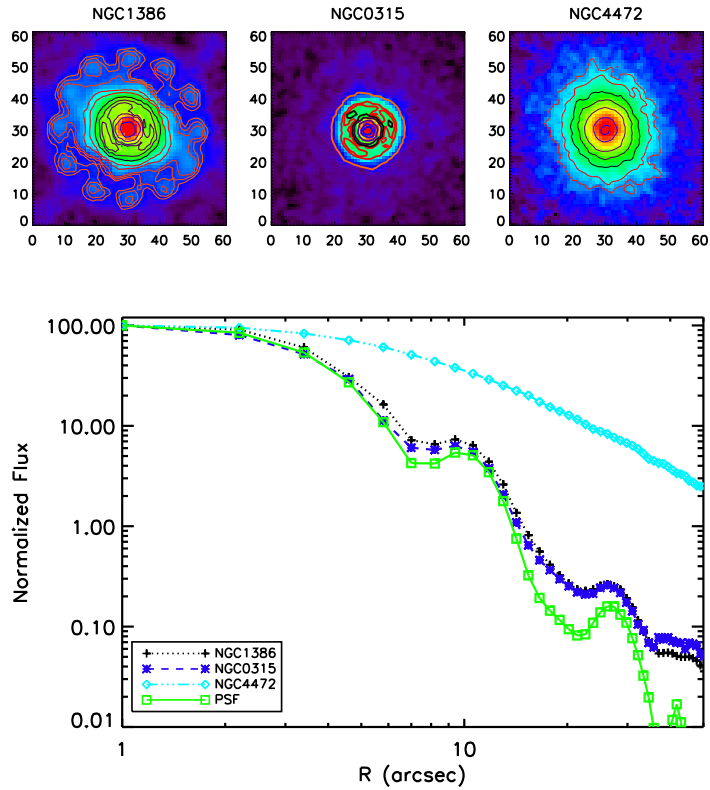}
\vskip2.7in
\caption{
Upper panels: 24$\mu$m images of NGC 1398 and 315 showing Airy
diffraction rings and NGC 4472 with diffraction-free image. The colors
and contours are qualitative only, intended to illustrate
non-axisymmetric features in the diffraction pattern.  Lower panel:
Azimuthally averaged radial diffraction fluxes at 24$\mu$m for the
three galaxies imaged above: NGC 1386 (black dotted line); NGC 0315
(blue dashed line); NGC 4472 (light blue long dash-triple dotted
line).  For comparison we plot the point response function of the
Spitzer telescope at 24$\mu$m (green solid line). All fluxes are
normalized to 100 (in arbitrary units) at one arcsecond.
}
\label{fig2}
\end{figure}

\clearpage

\begin{figure}
\centering
\includegraphics[width=7.0in,scale=1.0,angle=0]{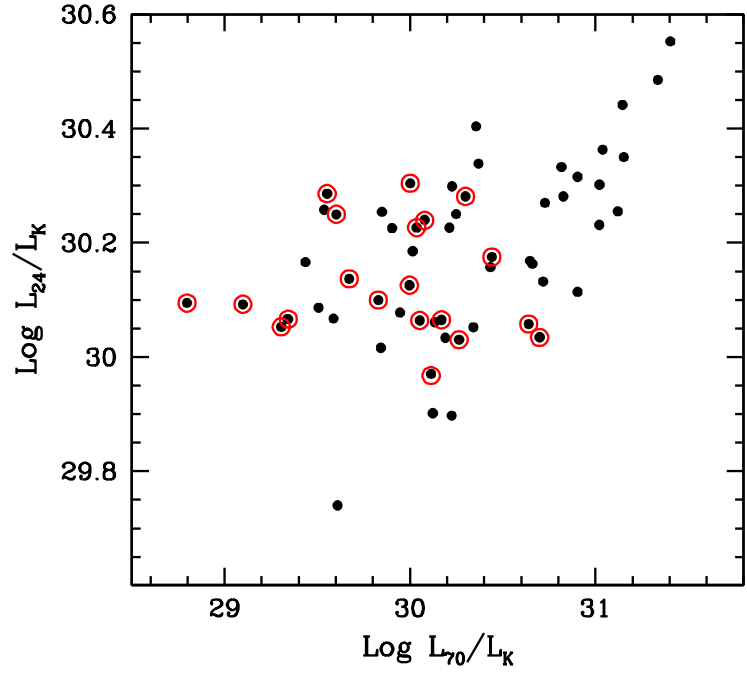}
\vskip.7in
\caption{
Expanded view of E and E-S0 galaxies (with $T < -2.7$) at the bottom
of the banana plot, omitting AGN galaxies with significant
point source emission at 24$\mu$m.  Red circles identify a subset of these
galaxies that are also in the sample of Trager et al. (2000a, b)
which have been observed at both 24 and 70$\mu$m but omitting two
that are identified as AGNs in Table 1, NGC 0315 and NGC 4261.
}
\label{fig3}
\end{figure}

\clearpage

\begin{figure}
\centering
\includegraphics[width=4.5in,scale=1.0,angle=0]{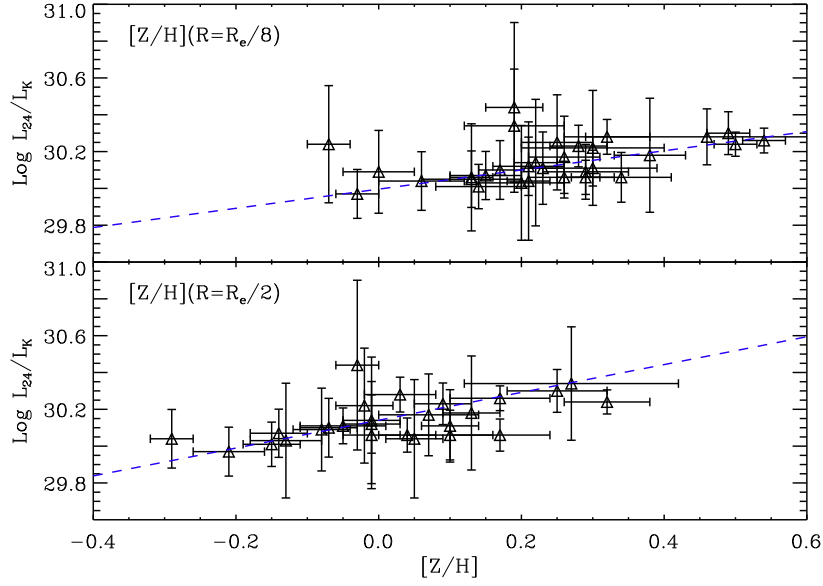}
\vskip.7in
\caption{
Correlation between $L_{24}/L_K$ and stellar metallicity $[Z/H]$ for the
Trager et al. (2000a, b) sample of E galaxies. Upper and lower panels show
the linear relation with metallicities values derived in the $R_e/8$
and $R_e/2$ aperture respectively where $R_e$ 
is the effective radius in the $B$-band.
Two galaxies identified as AGNs in Table 1 (NGC 315 and NGC 4261)
are not included in either panel.
In the lower panel 
we also exclude the outlying (and very uncertain)
value $[Z/H] = 1.13$ for NGC 720.
}
\label{fig4}
\end{figure}

\clearpage

\begin{figure}
\centering
\includegraphics[width=4.5in,scale=1.0,angle=0]{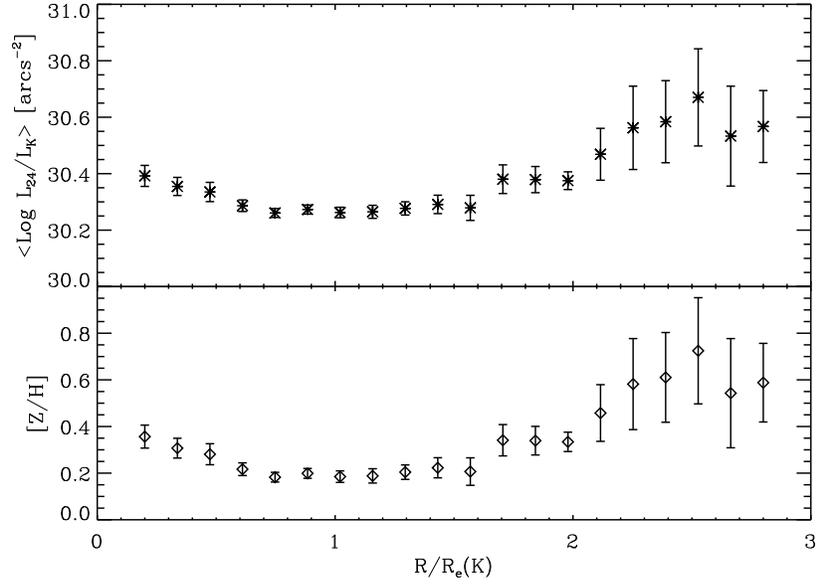}
\vskip.7in
\caption{
Upper panel: Variation of Trager sample-averaged
radial color profile (in surface brightness units)
with galaxy radius normalized to the K-band half
light radius $R_e(K)$.
Lower panel: Variation of Trager sample-averaged
radial stellar metallicity profiles found
using the correlation in Figure 4 derived for the $R_e/2$ aperture.
}
\label{fig5}
\end{figure}

\clearpage

\begin{figure}
\centering
\includegraphics[width=7.in,scale=1.0,angle=0]{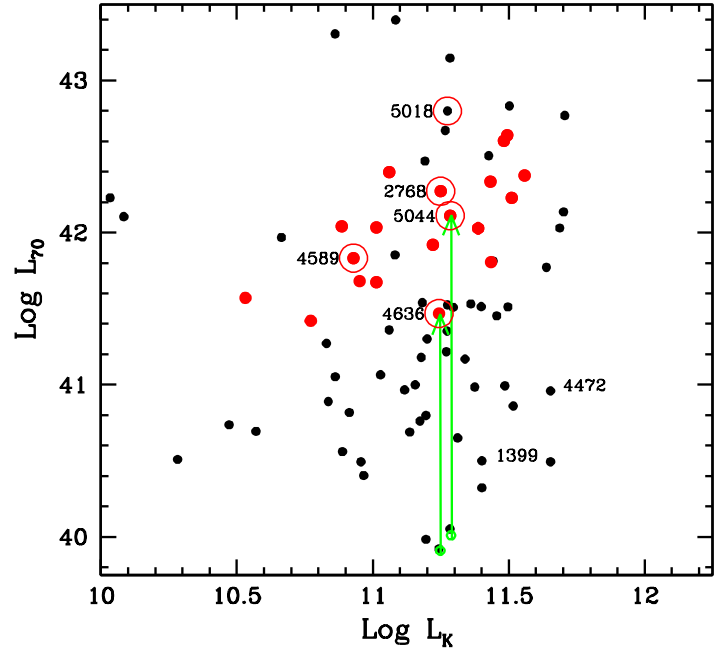}
\vskip.7in
\caption{
Plot of $L_{70}$ vs. $L_K$ for elliptical galaxies
($T < -3.5$) in the TMB09b sample.
%The dashed green horizontal line indicates an approximate
%locus of the intrinsic $L_{70}$ for
%$\log L_K \approx 11-11.5$ when no extended excess
%cold dust is present.
Galaxies marked in red all have excess extended
dust and the large red circles indicate galaxies
in which the spatial extension has been
observed at 70$\mu$m.
Vertical green arrows show the approximate estimated
increase in $L_{70}$ due to excess cold dust
in NGC 5044 and 4636.
}
\label{fig6}
\end{figure}

\clearpage

\begin{figure}
\centering
\includegraphics[width=7.in,scale=1.0,angle=0]{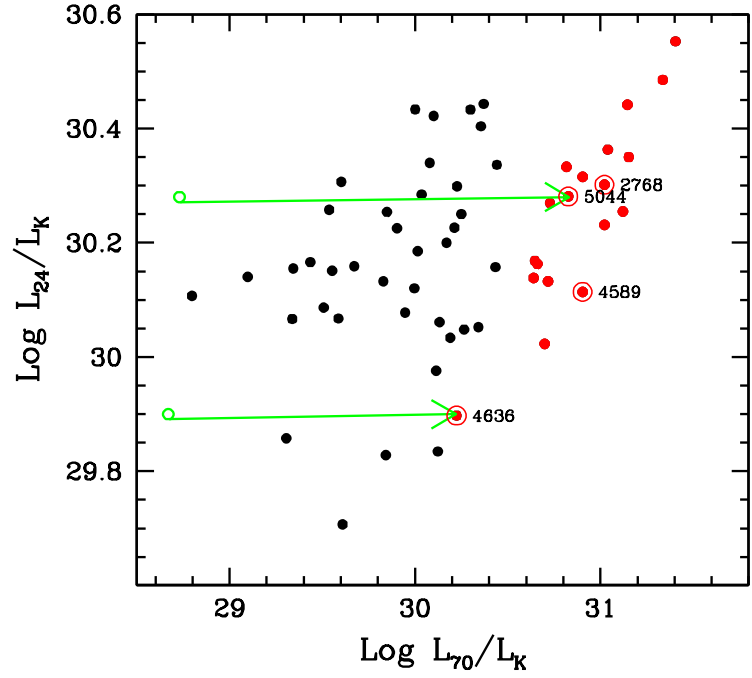}
\vskip.7in
\caption{
Infrared color-color plot as in Figure 3 showing
only E and E-S0 galaxies with $T \le -3.5$.
Galaxies in red and circled in red
are identical with those in
Figure 6, but now appear along the
right boundary of the lower banana.
As in Figure 6 the horizontal green arrows
show the trajectories of NGC 5044 and 4636
when extended cold dust is added to 
70$\mu$m emission 
produced only by local AGB stellar mass loss.
}
\label{fig7}
\end{figure}

\clearpage

\begin{figure}
\centering
\includegraphics[width=7.in,scale=1.0,angle=0]{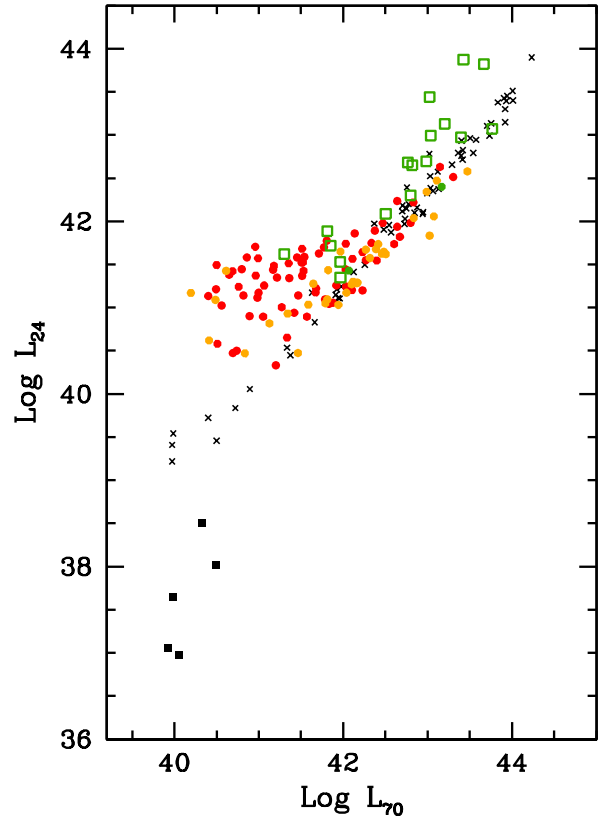}
\vskip.7in
\caption{
Plot of $L_{24}$ against $L_{70}$ as in 
the bottom panel of Figure 1
but without the $L_K$ normalization. 
Symbols are identical to those in that panel.
%Filled squares show positions of five elliptical 
%galaxies based on computations of steady state 
%emission from dust distributed smoothly in the 
%hot interstellar gas.
Filled squares show computed steady-state  
emission from diffusely distributed interstellar dust 
in five representative elliptical galaxies 
(NGC 4472, 1404 4636, 1399 \& 5044 
in order of decreasing $L_{24}$). 
The star formation locus of SINGS galaxies 
(crosses) is $L_{24} \propto L_{70}^{1.08}$.
}
\label{fig8}
\end{figure}

\end{document}